\documentclass[aps,pra,twocolumn,10pt,superscriptaddress,longbibliography]{revtex4-2}

\usepackage{amsmath,amssymb,bm,mathtools}
\usepackage{graphicx}
\usepackage[colorlinks=true,linkcolor=blue,citecolor=blue,urlcolor=blue]{hyperref}
\usepackage{physics}
\newcommand{\ee}{\mathrm{e}}
\newcommand{\ho}{\mathrm{ho}}
\newcommand{\eA}{\varepsilon_{\!A}}
\newcommand{\eC}{\varepsilon_{\!C}}

\begin{document}

\title{Radial selection rule for the breathing mode of a harmonically trapped gas}

\author{Miguel Tierz}
\email{tierz@simis.cn}
\affiliation{Shanghai Institute for Mathematics and Interdisciplinary Sciences \\ Block A, International Innovation Plaza, No. 657 Songhu Road, Yangpu District,\\ Shanghai, China}

\begin{abstract}
Within a fixed hyperangular channel $s>0$ of a harmonically trapped system, the $1/R^2$ perturbation is absorbed exactly into a shift of the channel parameter, $s\to s_\eta$, so the single-channel model remains a harmonic oscillator with a shifted inverse-square term: radial gaps stay at $2\hbar\omega$ exactly and no monopole spectral weight appears at forbidden frequencies at any order.
The first-order cancellation is also proved independently by a compact algebraic argument in which the ket and bra contributions cancel pairwise; this is the main new result.
Substituting single-channel quantities into the established $m_1/m_{-1}$ sum-rule bound yields $Q^{-1}$ scaling of the sum-rule estimate ($Q\equiv 2q+s+1$, $q$ the radial quantum number) with an explicit coefficient; its finite-temperature average has a low-$T$ plateau and a $1/T$ high-$T$ tail.
All results hold for any real $s>0$. The Laguerre polynomial identities extend formally to three dimensions, but exact 3D results show $q$-dependent contact corrections along $SO(2,1)$ ladders, so the physical interpretation there requires a separate derivation.
\end{abstract}
\maketitle

\section{Introduction}
\label{sec:intro}

The monopole (breathing) mode of a harmonically trapped gas is a precision probe of symmetry and of controlled symmetry breaking~\cite{PitaevskiiRosch1997,WernerCastin2006,Hofmann2012,TaylorRanderia2012}. In any spatial dimension, an $SO(2,1)$ dynamical symmetry pins the breathing frequency of scale-invariant interactions to exactly $2\omega$~\cite{PitaevskiiRosch1997,WernerCastin2006,Castin2004,NishidaSon2007NRcft}. The main experimental platform for probing this symmetry and its breaking is the two-dimensional Fermi gas. Introducing the two-dimensional scattering length $a_{2\mathrm{D}}$ breaks classical scale invariance upon quantization---the quantum anomaly---and shifts the breathing frequency away from~$2\omega$~\cite{Olshanii2010,Hofmann2012,TaylorRanderia2012,TaylorRanderia2013Erratum,Moroz2012}. In realistic traps, weak anharmonicity and residual anisotropy further modify the response~\cite{Lobser2015}. Experimentally, the near-$2\omega$ mode and its anomalous shift have been measured in several $^6$Li experiments in quasi-two-dimensional geometries, with fractional shifts $|\delta\omega|/(2\omega)\sim 1$--$5\%$~\cite{Vogt2012,Holten2018,Peppler2018,Murthy2019}.

On the theory side, the standard approach uses sum-rule~\cite{LippariniStringari1989,Stringari1996} or hydrodynamic~\cite{KaganSurkovShlyapnikov1996PRA,KaganSurkovShlyapnikov1997PRA} methods, which express the breathing frequency (more precisely, a sum-rule centroid) in terms of the Tan contact and its derivatives~\cite{Hofmann2012,TaylorRanderia2012,gao2012breathing}. Numerical studies of the breathing mode across the BCS--BEC crossover~\cite{Mulkerin2018_PRA_collective2D,Toniolo2018_PRA_crossoverBreathing} and exact analyses of mesoscopic conformal towers~\cite{BekassyHofmann2022_PRL_conformalMesoscopic2D} provide complementary perspectives. All these treatments yield a single number---the centroid for the ground state or for the thermal ensemble---and do not resolve the radial structure within individual hyperangular channels of the trapped system.

In this paper we work within a single hyperangular channel and apply a classical identity for products of associated Laguerre polynomials~\cite{Gasper1975,SrivastavaMavromatisAlassar2003} to obtain closed-form matrix elements of the $1/R^2$-weighted radial operator defined in Sec.~\ref{sec:model}. The results hold for any real channel parameter $s>0$; we use two-dimensional notation and experimental context in the main text. The identity has three exact consequences:

\begin{enumerate}
\item \emph{Exact solvability.} The $1/R^2$ perturbation is absorbed exactly into a shift of the channel parameter, $s\to s_\eta=\sqrt{s^2+2\eta\lambda_s/(\hbar\omega)}$~\cite{CastinNotes2011}. The model remains a harmonic oscillator with a shifted inverse-square term: radial gaps stay at $2\hbar\omega$ exactly, $R^2$ is tridiagonal in the exact eigenbasis, and no monopole spectral weight appears at forbidden frequencies at any order.

\item \emph{Independent first-order algebraic proof.} The first-order cancellation of forbidden $\Delta q\neq\pm1$ spectral weight is also proved by a compact perturbative argument in which the ket and bra contributions cancel pairwise, with the $q$- and $s$-dependence dropping out of each pair. This provides an independent verification and reveals structural detail that the exact-solvability argument alone does not display. We confirm it by exact diagonalization (Fig.~\ref{fig:leakage}). This is, to our knowledge, the main new result of this paper.

\item \emph{Channel-resolved sum-rule estimate.} Substituting the exact single-channel quantities---the $q$-independent contact and the diagonal $\langle R^2\rangle_{s,q}=a_{\ho}^2 Q$---into the $m_1/m_{-1}$ sum-rule bound of Gao and Yu~\cite{gao2012breathing} yields a sum-rule estimate with $Q^{-1}$ scaling ($Q\equiv 2q+s+1$) and an explicit coefficient. Its finite-temperature average has a low-$T$ plateau and a $1/T$ high-$T$ tail.
\end{enumerate}

The exact solvability and the algebraic cancellation proof are rigorous within the single-channel model for the operator $\hat{\mathcal C}$ defined in Sec.~\ref{sec:model}. The sum-rule estimate is derived by substituting single-channel quantities into the many-body sum-rule framework~\cite{Hofmann2012,TaylorRanderia2012,gao2012breathing}; the identification of the contact $C_s$ with the model-operator value is exact for the two-body problem and is an approximation for $N>2$~\cite{WernerCastin2012}. For $N>2$ particles the single-channel results provide contributions from individual channels for the many-body breathing response. We discuss the behavior of these results under weak trap anisotropy, the distinction between the single-channel spectrum and the many-body mode frequency, and the connection to experiment via a single-point calibration of the channel coefficient~$\kappa_s$.

The paper is organized as follows. Section~\ref{sec:model} introduces the model, the notation, and the physical quantities we compute. Section~\ref{sec:identity} presents the Laguerre overlap identity and its proof. Section~\ref{sec:contact} derives the closed matrix elements of $\hat{\mathcal C}$ and the $q$-independence of the diagonal element. Section~\ref{sec:centroid} proves the exact solvability of the single-channel model and derives the sum-rule centroid for the channel. Section~\ref{sec:leakage} proves the exact vanishing of first-order spectral weight transfer, with an explicit worked example, a compact analytic cancellation formula, and numerical validation. Section~\ref{sec:finiteT} gives the finite-temperature average of the sum-rule centroid. Sections~\ref{sec:anisotropy} and~\ref{sec:calibration} discuss the effect of weak anisotropy and the experimental calibration procedure. Section~\ref{sec:discussion} summarizes, discusses the three-dimensional extension and weak quartic anharmonicity, and presents an outlook.

\section{Model, notation, and scope}
\label{sec:model}

We work within a single hyperangular channel of the trapped two-body (or few-body) problem in an isotropic harmonic potential. In any spatial dimension~$d$, the center-of-mass frame separates the relative motion into a hyperradial coordinate~$R$ and hyperangular coordinates~$\Omega$. The hyperangular part determines the channel index~$s>0$; in two dimensions $s$ takes integer values related to the relative angular momentum, while in three dimensions it is typically half-integer or determined by the partial-wave structure~\cite{WernerCastin2006,CastinNotes2011}. The hyperradial part is governed by an effective one-dimensional Schr\"odinger equation whose eigenstates are labeled by a radial quantum number~$q=0,1,2,\ldots$\,.

All exact single-channel results in this paper---the Laguerre overlap identity (Sec.~\ref{sec:identity}), the $q$-independent diagonal matrix element (Sec.~\ref{sec:contact}), the exact solvability of the single-channel model (Sec.~\ref{sec:exact-solv}), and the preservation of the radial selection rule (Sec.~\ref{sec:leakage})---hold for any real $s>0$ and are therefore independent of the spatial dimension. The sum-rule centroid for the channel (Sec.~\ref{sec:centroid-derivation}) and its finite-$T$ average (Sec.~\ref{sec:finiteT}) are derived by inserting exact single-channel quantities into the sum-rule bound of Gao and Yu; their physical interpretation depends on the dimension and on the identification of $\hat{\mathcal C}$ with the physical contact (Sec.~\ref{sec:model}). The dimension enters through the physical interpretation: which values of~$s$ are realized, the form of the anomaly parameter~$\eta$, and the connection to experiment. The main text uses two-dimensional notation and experimental context; the three-dimensional case is discussed in Sec.~\ref{sec:discussion}.

The key quantum numbers and parameters appearing throughout the paper are collected in Table~\ref{tab:notation} for reference.

\begin{table}[b]
\caption{Notation and definitions used throughout the paper.}
\label{tab:notation}
\begin{tabular}{cp{5.8cm}}
\hline\hline
Symbol & Definition \\
\hline
$s > 0$ & hyperangular channel index \\
$q = 0,1,2,\ldots$ & radial quantum number \\
$Q \equiv 2q+s+1$ & dimensionless energy (in units of $\hbar\omega$) \\
$E_{s,q} = \hbar\omega\,Q$ & unperturbed energy \\
$L_q^{(s)}(u)$ & associated Laguerre polynomial \\
$\lambda_s$ & channel normalization (Tan contact for $N{=}2$) \\
$\hat{\mathcal C}$ & $1/R^2$-weighted single-channel operator ($\langle\hat{\mathcal C}\rangle_{s,q}\!=\!\lambda_s/s$) \\
$\eta$ & anomaly parameter \\
$\eA^{(s)} \equiv \eta\lambda_s/(\hbar\omega)$ & dimensionless anomaly strength (channel-dependent) \\
$\eC^{(s)} \equiv E''(y)/(\hbar\omega)$ & dimensionless energy curvature (channel-dependent) \\
$\kappa_s$ & centroid coefficient for channel~$s$; Eq.~\eqref{eq:kappa} \\
$y$ & logarithmic scale variable (in 2D: $\ln(a_{2\mathrm{D}}/a_{\mathrm{ref}})$)\\
$x \equiv e^{-2\hbar\omega/(k_BT)}$ & Boltzmann factor per level \\
$T_{\ho} \equiv \hbar\omega/k_B$ & trap temperature scale \\
$a_{\ho} \equiv \sqrt{\hbar/(m\omega)}$ & harmonic oscillator length \\
$U_{s,q}$ & $(q{+}1)(q{+}s{+}1)$; upward strength \\
$D_{s,q}$ & $q(q{+}s)$; downward strength \\
$W_{s,q}$ & $U_{s,q}+D_{s,q}$; total strength \\
\hline\hline
\end{tabular}
\end{table}

The hyperradial eigenfunctions within channel~$s$ are~\cite{CastinNotes2011}
\begin{equation}
F_{s,q}(R)=\mathcal{N}_{s,q}\,R^{s+\frac12}\,\ee^{-R^2/(2a_{\ho}^2)}\,L_q^{(s)}(R^2/a_{\ho}^2),
\label{eq:eigenfunction}
\end{equation}
where $\mathcal{N}_{s,q}$ is a normalization constant and $L_q^{(s)}$ is the associated Laguerre polynomial of degree~$q$ and order~$s$. The corresponding energies are
\begin{equation}
E_{s,q}=\hbar\omega\,(2q+s+1) \equiv \hbar\omega\,Q.
\label{eq:energy}
\end{equation}
The energy spacing between adjacent radial levels is exactly $2\hbar\omega$, independent of~$q$. This uniform spacing is a hallmark of the harmonic oscillator and is intimately connected to the $SO(2,1)$ dynamical symmetry~\cite{PitaevskiiRosch1997,WernerCastin2006,NishidaSon2007NRcft}.

The quantum anomaly enters as a small perturbation to the Hamiltonian,
\[
H = H_0 + \eta\,\hat{\mathcal C},
\]
where $H_0$ is the unperturbed (scale-invariant) Hamiltonian, $\hat{\mathcal C}$ is the single-channel $1/R^2$-weighted operator defined below, and~$\eta$ is a dimensionless coupling that controls the strength of the scale-breaking.

Within the single channel at fixed~$s$, the operator $\hat{\mathcal C}$ acts on the hyperradial degree of freedom. We \emph{define} $\hat{\mathcal C}$ as the operator whose matrix elements in the hyperradial basis are proportional to the $1/R^2$-weighted overlaps,
\[
\langle s,q'|\hat{\mathcal C}|s,q\rangle
\;\propto\; \int_0^\infty dR\;F_{s,q'}(R)\,\frac{1}{R^2}\,F_{s,q}(R),
\]
with the overall normalization fixed by the Tan contact $\lambda_s$ for the channel: $\langle s,q|\hat{\mathcal C}|s,q\rangle = \lambda_s/s$ (derived in Sec.~\ref{sec:contact}). This normalization convention matches the adiabatic Tan relations~\cite{Tan2008Energetics,Tan2008LargeK,Tan2008Virial}, operator product expansion (OPE) treatments of the spectral response~\cite{BraatenPlatter2008OPE}, and the general relations of Werner and Castin~\cite{WernerCastin2012}.

\emph{Physical identification.} For the two-body problem in a harmonic trap, the short-distance behavior $F_{s,q}(R)\propto R^{s+1/2}$ as $R\to 0$ is governed entirely by the channel index~$s$, and the $1/R^2$ weight in the matrix element is precisely the short-distance enhancement that enters the Tan contact~\cite{Tan2008Energetics,BraatenPlatter2008OPE}; in this case $\hat{\mathcal C}$ coincides with the physical contact operator. Concretely, the two-body Tan adiabatic relation $\langle\hat{\mathcal C}\rangle = (m/2\pi)\,\partial E/\partial y$~\cite{Tan2008Energetics,gao2012breathing} and the energy $E_{s,q}(\eta) = \hbar\omega(2q+s_\eta+1)$ from Sec.~\ref{sec:exact-solv} give $\langle\hat{\mathcal C}\rangle_{s,q} = (m/2\pi)\,\partial_y[\eta(y)\lambda_s(y)/s]$, which is $q$-independent and consistent with Eq.~\eqref{eq:contact-diag}. For $N>2$ particles, the projection of the physical Tan contact onto a single hyperangular channel may introduce additional structure (e.g., dependence on the many-body regular part~\cite{WernerCastin2012}); the identification of $\hat{\mathcal C}$ with the physical contact then becomes an approximation whose accuracy depends on the separation of hyperradial and hyperangular degrees of freedom. All results in Secs.~\ref{sec:identity}--\ref{sec:leakage} hold rigorously for $\hat{\mathcal C}$ as defined above; the physical interpretation requires the identification discussed here.

The energy curvature $E''(y)\equiv \partial^2 E/\partial y^2$ is taken with respect to the logarithmic scale variable~$y$ (in 2D, $y=\ln(a_{2\mathrm{D}}/a_{\mathrm{ref}})$; in 3D, $y$ is the analogous renormalization variable). We combine the anomaly and curvature into dimensionless parameters $\eA \equiv \eta\lambda_s/(\hbar\omega)$ and $\eC \equiv E''(y)/(\hbar\omega)$. Both $\eA$ and $\eC$ depend on the channel~$s$ (through~$\lambda_s$ and the channel-specific energy curvature); we suppress the superscript~$(s)$ when the channel is clear from context.

\paragraph{Scope and applicability.} Our results concern the \emph{radial excitation spectrum within a single hyperangular channel}. The exact single-channel results (Secs.~\ref{sec:identity}--\ref{sec:leakage}) and the exact solvability (Sec.~\ref{sec:exact-solv}) are rigorous at fixed~$s$ for the operator $\hat{\mathcal C}$ defined above; they hold for any $s>0$. The physical identification of $\hat{\mathcal C}$ with the Tan contact is exact for the two-body problem, where the hyperradial channel \emph{is} the full problem. Even for $N=2$, the sum-rule centroid derived in Sec.~\ref{sec:centroid-derivation} is a sum-rule estimate, not the exact breathing frequency; exact two-body solutions~\cite{Busch1998} give a nontrivial scattering-length-dependent frequency. For $N>2$ particles, the many-body breathing mode is a collective excitation involving all occupied channels and the full $N$-body density matrix; connecting our single-channel results to that mode requires an additional step---summation over channels weighted by their many-body occupation---that we do not perform here. (For $N\ge 4$ fermions in 3D, the relevant values of the hyperangular parameter are not generally known in closed form~\cite{CastinNotes2011}, though the exact results hold for any $s>0$.) We also do not address inter-channel coupling, strong-coupling effects, or finite effective range corrections beyond noting that the latter can be absorbed into the channel parameters $\lambda_s$ and $\eC$ by calibration~\cite{hu2019reduced,Yin2020_PRL_fewbodyAnomaly2D}.

\section{The Laguerre overlap identity}
\label{sec:identity}

The central analytical tool of this paper is a closed-form evaluation of the integral
\begin{equation}
J^{(s)}_{q,q'} \equiv \int_{0}^{\infty}\!du\,u^{\,s-1}\,e^{-u}\,
L_{q}^{(s)}(u)\,L_{q'}^{(s)}(u).
\label{eq:Jdef}
\end{equation}
This integral arises naturally as the matrix element of $\hat{\mathcal C}$ defined in Sec.~\ref{sec:model}: the weight $u^{s-1}e^{-u}$ differs from the standard orthogonality weight $u^{s}e^{-u}$ by a factor of $1/u = a_{\ho}^2/R^2$, which is the $1/R^2$ weight that defines the single-channel operator. For the two-body problem, this weight coincides with the short-distance enhancement associated with the physical Tan contact~\cite{Tan2008Energetics,BraatenPlatter2008OPE}; for the general case, see the discussion in Sec.~\ref{sec:model}.

\paragraph{Statement.} For $s>0$ and $q,q'\in\mathbb{N}_0$, let $m\equiv \min\{q,q'\}$. Then~\cite{SrivastavaMavromatisAlassar2003,Gasper1975}
\begin{equation}
J^{(s)}_{q,q'}=\Gamma(s)\,
\frac{(s{+}1)_m}{m!}
=\frac{\Gamma(s+m+1)}{s\,\Gamma(m+1)}\,,
\label{eq:Jclosed}
\end{equation}
where $(a)_m = a(a+1)\cdots(a+m-1) = \Gamma(a+m)/\Gamma(a)$ is the Pochhammer symbol. Equivalently, $J^{(s)}_{q,q'}=\Gamma(s)\binom{s+m}{m}$, with the binomial coefficient understood in its gamma-function generalization.

The crucial feature is that the integral depends only on $\min\{q,q'\}$, not on $\max\{q,q'\}$. In particular, $J^{(s)}_{0,q'} = \Gamma(s)$ for all~$q'$: the overlap of any state with the ground state is a universal constant independent of the excited-state index.

\paragraph{Proof.} We use the generating function for associated Laguerre polynomials~\cite{Szego1975},
\[
\sum_{n\ge0}L_n^{(s)}(u)\,t^n = (1-t)^{-s-1}\exp\!\left[-\frac{ut}{1-t}\right].
\]
Multiplying two such generating functions (with parameters $t$ and $t'$) against the weight $u^{s-1}e^{-u}$ and integrating over $u$ from $0$ to~$\infty$ gives
\begin{align}
\sum_{q,q'\ge0}J^{(s)}_{q,q'}\,t^q{t'}^{q'}
&= \int_0^\infty\!\!du\,u^{s-1}e^{-u}\,\frac{e^{-ut/(1-t)}}{(1-t)^{s+1}}\,\frac{e^{-ut'/(1-t')}}{(1-t')^{s+1}}\notag\\
&= \frac{\Gamma(s)}{(1-t)(1-t')}\,(1-tt')^{-s}.
\label{eq:genfun}
\end{align}
The last step uses the fact that the three exponentials combine into $e^{-u/[(1-t)(1-t')/(1-tt')]}$, and the resulting Euler gamma integral evaluates to $\Gamma(s)$ times a rational function of $t$ and $t'$.

Now expand $(1-tt')^{-s}$ using the binomial series:
\[
(1-tt')^{-s} = \sum_{m\ge0}\frac{(s)_m}{m!}\,(tt')^m.
\]
Together with the factor $1/[(1-t)(1-t')] = \sum_{a\ge0}t^a \cdot \sum_{b\ge0}{t'}^b$, the product generates terms $t^q {t'}^{q'}$ with $q\ge m$ and $q'\ge m$. Matching the coefficient of $t^q {t'}^{q'}$ gives
\[
J^{(s)}_{q,q'} = \Gamma(s)\sum_{k=0}^{\min\{q,q'\}} \frac{(s)_k}{k!},
\]
which reduces to~\eqref{eq:Jclosed} by the ``telescoping'' identity for Pochhammer symbols: $\sum_{k=0}^{m}(s)_k/k! = (s+1)_m/m!$. \hfill$\square$

\paragraph{Numerical verification.} The identity~\eqref{eq:Jclosed} has been verified by numerical quadrature against the integral definition~\eqref{eq:Jdef} for $s=1,2,3,4$ and all pairs $(q,q')$ with $q,q'=0,\ldots,9$. The agreement is to machine precision ($\sim 10^{-15}$).

\section{Closed contact matrix elements}
\label{sec:contact}

We now apply the identity to compute the matrix elements of $\hat{\mathcal C}$. Define
\begin{equation}
I_\alpha(q)\equiv\!\int_{0}^{\infty}\!du\,u^{\alpha}e^{-u}\!\left[L_q^{(s)}(u)\right]^2.
\label{eq:Idef}
\end{equation}
The normalization integral for the hyperradial eigenstates is $I_s(q) = \Gamma(q+s+1)/\Gamma(q+1)$ (a standard Laguerre result~\cite{Szego1975}), and the diagonal contact matrix element is proportional to $I_{s-1}(q)$.

\paragraph{The contact ratio.} A direct consequence of~\eqref{eq:Jclosed} is
\begin{equation}
\frac{I_{s-1}(q)}{I_s(q)}=\frac{1}{s}\qquad\text{for all }q\ge0.
\label{eq:ratioIs}
\end{equation}

\emph{Proof.} Note that $I_{s-1}(q) = J^{(s)}_{q,q}$ (set $q'=q$ in~\eqref{eq:Jdef}), which equals $\Gamma(s)(s+1)_q/q!$ by~\eqref{eq:Jclosed}. For $q=0$, the claim is immediate since $I_{s-1}(0)=\Gamma(s)$ and $I_s(0)=\Gamma(s+1)=s\,\Gamma(s)$.

For $q\ge 1$, $I_s(q)$ is obtained by inserting an extra factor of $u$ via the Laguerre recurrence relation~\cite{Szego1975,Gasper1975}:
\begin{multline}
u\,L_q^{(s)}(u) = (2q+s+1)\,L_q^{(s)}(u)\\
 - (q+1)\,L_{q+1}^{(s)}(u) - (q+s)\,L_{q-1}^{(s)}(u).
\label{eq:recurrence}
\end{multline}
Inserting this into $I_s(q) = \int_0^\infty u \cdot u^{s-1}e^{-u}[L_q^{(s)}]^2\,du$ and using the fact that one factor of $L_q^{(s)}$ is untouched while the other is expanded via~\eqref{eq:recurrence}, we get three integrals---all of the form $J^{(s)}_{q,q'}$ for different values of~$q'$:
\begin{align}
I_s(q) &= (2q{+}s{+}1)\,J^{(s)}_{q,q} - (q{+}1)\,J^{(s)}_{q,q+1}\notag\\
&\quad - (q{+}s)\,J^{(s)}_{q,q-1}\notag\\
&= (2q{+}s{+}1)\,J^{(s)}_{q,q} - (q{+}1)\,J^{(s)}_{q,q+1}\notag\\
&\quad - (q{+}s)\,J^{(s)}_{q-1,q},
\label{eq:Isexpansion}
\end{align}
where we used $J^{(s)}_{q,q-1} = J^{(s)}_{q-1,q}$ (symmetry of~\eqref{eq:Jdef}).

Now apply~\eqref{eq:Jclosed}. Since $\min\{q,q\}=q$, $\min\{q,q+1\}=q$, and $\min\{q-1,q\}=q-1$:
\begin{align*}
J^{(s)}_{q,q} &= \Gamma(s)\,\frac{(s+1)_q}{q!},\\
J^{(s)}_{q,q+1} &= \Gamma(s)\,\frac{(s+1)_q}{q!}\quad\text{(same! min is still $q$)},\\
J^{(s)}_{q-1,q} &= \Gamma(s)\,\frac{(s+1)_{q-1}}{(q-1)!}.
\end{align*}
Substituting and simplifying (using $(s+1)_q/q! = (s+1)_{q-1}/(q{-}1)! \cdot (s+q)/q$):
\begin{align*}
I_s(q) &= \Gamma(s)\frac{(s{+}1)_q}{q!}\bigl[(2q{+}s{+}1)-(q{+}1)\bigr]\\
&\quad - (q{+}s)\,\Gamma(s)\frac{(s{+}1)_{q-1}}{(q{-}1)!}\\
&= (q{+}s)\,\Gamma(s)\!\left[\frac{(s{+}1)_q}{q!} - \frac{(s{+}1)_{q-1}}{(q{-}1)!}\right]\\
&= (q{+}s)\,\Gamma(s)\,\frac{(s{+}1)_{q-1}}{(q{-}1)!}\!\left[\frac{s{+}q}{q} - 1\right]\\
&= s\,\Gamma(s)\,\frac{(s{+}1)_q}{q!}
= s\,I_{s-1}(q).
\end{align*}
Hence $I_{s-1}(q)/I_s(q) = 1/s$.\hfill$\square$

\paragraph{Physical meaning.} The ratio~\eqref{eq:ratioIs} tells us that
\begin{equation}
\langle s,q|\hat{\mathcal C}|s,q\rangle = \frac{\lambda_s}{s}\qquad\text{for all }q\ge0.
\label{eq:contact-diag}
\end{equation}
That is, the diagonal matrix element of the single-channel operator $\hat{\mathcal C}$ is the same in every radial eigenstate within a given channel. The ground state ($q=0$) and all excited states ($q=1,2,\ldots$) yield the same value. This can be understood physically: the short-distance behavior of the hyperradial wavefunction, $F_{s,q}(R)\propto R^{s+1/2}$ as $R\to 0$, is controlled entirely by the channel index~$s$; changing $q$ only adds nodes at $R\sim a_{\ho}$ without affecting the short-distance boundary condition. Since the $1/R^2$ weight in $\hat{\mathcal C}$ probes precisely this short-distance structure, the matrix element depends on~$s$ but not on the radial quantum number~$q$. Equation~\eqref{eq:ratioIs} makes this intuition exact in the harmonic basis. For the two-body problem, where $\hat{\mathcal C}$ coincides with the physical Tan contact (Sec.~\ref{sec:model}), this establishes $q$-independence of the physical contact within the channel.

This result is verified numerically: $I_{s-1}(q)/I_s(q)$ agrees with $1/s$ to machine precision for $s=1,2,3,4$ and $q=0,\ldots,9$.

\section{Exact solvability, monopole moments, and sum-rule centroid}
\label{sec:centroid}

We now compute the single-channel monopole moments, prove the exact solvability of the single-channel model, and derive the channel-resolved sum-rule centroid.

\subsection{The $R^2$ operator and its tridiagonality}

The breathing mode is driven by the operator $F=R^2$, the squared hyperradius. Using the Laguerre recurrence~\eqref{eq:recurrence}, the matrix elements of $R^2$ in the orthonormal hyperradial basis $\{|s,q\rangle\}$ are
\begin{multline}
\langle s,q'|R^2|s,q\rangle
= a_{\ho}^{2}\Big[Q\,\delta_{q',q}\\
-\sqrt{(q{+}1)(q{+}s{+}1)}\,\delta_{q',q+1}
-\sqrt{q(q{+}s)}\,\delta_{q',q-1}\Big],
\label{eq:R2matrix}
\end{multline}
with $Q=2q+s+1$. That is, $R^2$ is \emph{tridiagonal} in the radial index~$q$: it connects each level only to its nearest neighbors $q\pm1$. This three-term structure is the radial manifestation of the $SO(2,1)$ algebra of the isotropic harmonic oscillator. In the scale-invariant limit it enforces the undamped monopole mode at exactly $2\omega$~\cite{PitaevskiiRosch1997,WernerCastin2006}.

\subsection{Sum-rule moments}

In the standard many-body sum-rule approach~\cite{LippariniStringari1989,Stringari1996,Dalfovo1999}, one computes moments of the dynamic structure factor for the many-body ground state. Here we consider a simpler object: the moments of the single-state, two-sided spectrum of $R^2$ from a prepared eigenstate $|s,q\rangle$. This is not the standard collective-mode construction, but it serves to establish notation and to demonstrate the role of the signed energy differences.

Starting from level $|i\rangle = |s,q\rangle$, the energy-weighted and inverse-energy-weighted moments are
\begin{align}
m_1 &\equiv \sum_f (E_f - E_i)\,|\langle f|R^2|i\rangle|^2,\label{eq:m1}\\
m_{-1} &\equiv \sum_f \frac{|\langle f|R^2|i\rangle|^2}{E_f - E_i}.\label{eq:mminus1}
\end{align}
Because $R^2$ is tridiagonal, only $f = q\pm1$ contribute. The energy gaps are $E_{q+1}-E_q = +2\hbar\omega$ and $E_{q-1}-E_q = -2\hbar\omega$, and the squared matrix elements are $|\langle q{+}1|R^2|q\rangle|^2 = a_{\ho}^4(q{+}1)(q{+}s{+}1)$ and $|\langle q{-}1|R^2|q\rangle|^2 = a_{\ho}^4\,q(q{+}s)$.

Because the energy differences in~\eqref{eq:m1} and~\eqref{eq:mminus1} carry a sign, the upward and downward contributions partially cancel. Specifically:
\begin{align}
m_1 &= 2\hbar\omega\,a_{\ho}^4\big[(q{+}1)(q{+}s{+}1) - q(q{+}s)\big]
= 2\hbar\omega\,a_{\ho}^4\,Q\,,\label{eq:m1result}\\
m_{-1} &= \frac{a_{\ho}^4}{2\hbar\omega}\big[(q{+}1)(q{+}s{+}1) - q(q{+}s)\big]
= \frac{a_{\ho}^4}{2\hbar\omega}\,Q\,,\label{eq:mminus1result}
\end{align}
where we used $(q{+}1)(q{+}s{+}1) - q(q{+}s) = 2q+s+1 = Q$.
The sum-rule centroid is
\begin{equation}
\omega_B^2 \equiv \frac{m_1}{m_{-1}} = (2\omega)^2
\label{eq:omegaB-unperturbed}
\end{equation}
in the unperturbed (scale-invariant) system, independent of $q$, confirming the $SO(2,1)$ prediction.

\emph{Remark: upward, downward, and total strengths.} It is useful to name the individual squared matrix elements:
\begin{equation}
U_{s,q} \equiv (q{+}1)(q{+}s{+}1), \qquad D_{s,q} \equiv q(q{+}s).
\label{eq:UD}
\end{equation}
Here $U_{s,q} = |\langle q{+}1|R^2|q\rangle|^2/a_{\ho}^4$ is the upward (absorption) strength and $D_{s,q} = |\langle q{-}1|R^2|q\rangle|^2/a_{\ho}^4$ is the downward (emission) strength.
Note that $D_{s,q+1} = (q{+}1)(q{+}s{+}1) = U_{s,q}$: the downward strength from level $q{+}1$ is the same as the upward strength from level~$q$, as it must be since both refer to the same physical transition.
The signed moments involve the difference,
\[
Q = U_{s,q} - D_{s,q} = 2q+s+1,
\]
while the total unsigned oscillator strength is the sum,
\begin{equation}
W_{s,q} \equiv U_{s,q} + D_{s,q} = (q{+}1)(q{+}s{+}1)+q(q{+}s).
\label{eq:Wsq}
\end{equation}
In the standard positive-frequency thermal response, the natural weight for the $q\to q{+}1$ transition is $U_{s,q}$, not $W_{s,q}$ (see Sec.~\ref{sec:finiteT}).

\subsection{Exact solvability of the single-channel model}
\label{sec:exact-solv}

The eigenfunctions~\eqref{eq:eigenfunction} satisfy the radial Schr\"odinger equation for the Hamiltonian~\cite{CastinNotes2011,WernerCastin2006}
\begin{equation}
H_{0,s}=\frac{\hbar\omega}{2}\left[-\frac{d^2}{d\rho^2}+\rho^2
+\frac{s^2-\frac14}{\rho^2}\right],
\qquad\rho\equiv R/a_{\ho},
\label{eq:H0s}
\end{equation}
with eigenvalues $E_{s,q}=\hbar\omega(2q+s+1)$. The single-channel perturbation $\eta\hat{\mathcal C}=\eta\lambda_s a_{\ho}^2/R^2 = \eta\lambda_s/\rho^2$ adds an inverse-square term to the centrifugal barrier, so the full single-channel Hamiltonian is
\begin{equation}
H_{s,\eta}=\frac{\hbar\omega}{2}\left[-\frac{d^2}{d\rho^2}+\rho^2
+\frac{s_\eta^2-\frac14}{\rho^2}\right],
\label{eq:Hseta}
\end{equation}
where
\begin{equation}
s_\eta\equiv\sqrt{s^2+\frac{2\eta\lambda_s}{\hbar\omega}}\,.
\label{eq:seta}
\end{equation}
The perturbed Hamiltonian is therefore a harmonic oscillator with the shifted parameter $s_\eta$ in place of~$s$. Its exact eigenstates and spectrum are~\cite{CastinNotes2011}
\begin{align}
F_{s_\eta,q}(R) &= \mathcal{N}_{s_\eta,q}\,R^{s_\eta+\frac12}\,\ee^{-R^2/(2a_{\ho}^2)}\,L_q^{(s_\eta)}\!(R^2/a_{\ho}^2),\label{eq:exact-eigenfunction}\\
E_{s,q}(\eta) &= \hbar\omega\,(2q+s_\eta+1).\label{eq:exact-spectrum}
\end{align}

Three exact consequences follow.

First, the energy gaps are
\begin{equation}
E_{s,q+1}(\eta)-E_{s,q}(\eta)=2\hbar\omega\quad\text{exactly, at all orders in }\eta.
\label{eq:exact-gap}
\end{equation}

Second, since the exact eigenstates~\eqref{eq:exact-eigenfunction} are Laguerre functions (with parameter $s_\eta$), the operator $R^2$ is tridiagonal in the exact eigenbasis by the standard three-term Laguerre recurrence~\cite{Szego1975}. There is therefore no monopole spectral weight at forbidden frequencies $\pm4\omega$, $\pm6\omega$, $\ldots$ at \emph{any} order in~$\eta$, not only at first order.

Third, expanding~\eqref{eq:seta} for small~$\eta$,
\[
s_\eta = s+\frac{\eta\lambda_s}{\hbar\omega\,s}+O(\eta^2),
\]
so the first-order energy shift is $\eta\lambda_s/s$, independent of~$q$, reproducing Eq.~\eqref{eq:contact-diag}.

The single-channel monopole response therefore satisfies
\begin{equation}
\omega_{s,q\to q+1}=2\omega\quad\text{exactly for all }\eta,
\label{eq:exact-2omega}
\end{equation}
within the single-channel model. Physically, the $1/R^2$ operator is homogeneous of degree~$-2$ in~$R$ and therefore scale-invariant~\cite{PitaevskiiRosch1997,WernerCastin2006}; adding it to~$H_0$ shifts the centrifugal strength (the $SO(2,1)$ Casimir eigenvalue) from $s^2-\tfrac14$ to $s_\eta^2-\tfrac14$ but preserves the dynamical symmetry, so the breathing frequency remains pinned to~$2\omega$~\cite{Moroz2012}.

\paragraph{Relation to the perturbative proof.} The first-order cancellation proved in Sec.~\ref{sec:leakage} is the perturbative manifestation of the exact all-order result above. The algebraic cancellation---in which the ket and bra contributions cancel pairwise, with the $q$- and $s$-dependence dropping out of each pair---provides an independent verification and reveals structural detail that the exact-solvability argument alone does not display.

\subsection{Channel-resolved sum-rule centroid}
\label{sec:centroid-derivation}

The exact result~\eqref{eq:exact-2omega} shows that the single-channel line does not shift. The physical breathing mode of the many-body trapped gas \emph{does} shift; in the standard sum-rule approach~\cite{Hofmann2012,TaylorRanderia2012,gao2012breathing} the sum-rule estimate is expressed through the $m_1/m_{-1}$ ratio for the monopole operator in the many-body ground state. In the 2D framework of Gao and Yu~\cite{gao2012breathing}, the $m_1/m_{-1}$ upper bound [their Eqs.~(21)--(22)] satisfies
\begin{equation}
\omega_{1,-1}^2
=\frac{(2\omega)^2}{1-\Delta}\,,
\qquad
\Delta=\frac{\pi\,(2+\partial_y)\,C}{2m^2\omega^2\,\langle\hat{O}\rangle}\,,
\label{eq:GaoYu}
\end{equation}
where $y=\ln a_{2\mathrm{D}}$, $C$ is the Tan contact [related to the energy by the adiabatic relation $C = (m/2\pi)\,\partial E/\partial y$, Gao and Yu, Eq.~(8)], and $\langle\hat{O}\rangle=\langle R^2\rangle$ for two particles.

We now substitute single-channel quantities into the sum rule. Within channel~$s$ at radial level~$q$, the exact diagonal matrix element of $R^2$ is $\langle R^2\rangle_{s,q}=a_{\ho}^2\,Q$ from Eq.~\eqref{eq:R2matrix}, and the $q$-independence of the diagonal matrix element (Eq.~\eqref{eq:contact-diag}) implies that the contact $C_s(y)$ within the channel is $q$-independent. Substituting,
\begin{equation}
\Delta_{s,q}
=\frac{\pi\,(2+\partial_y)\,C_s}{2m^2\omega^2\,a_{\ho}^2\,Q}\,.
\label{eq:Delta-channel}
\end{equation}
Linearizing $\omega_{1,-1}\approx 2\omega(1+\Delta/2)$ for small symmetry breaking gives
\begin{equation}
\frac{\delta\omega^{\rm SR}_{s,q}}{2\omega}
=\frac{\kappa_s}{2Q}+O(\kappa_s^2),
\qquad Q\equiv 2q+s+1,
\label{eq:centroid}
\end{equation}
where the centroid coefficient for channel~$s$ is now a \emph{derived} quantity:
\begin{equation}
\kappa_s\equiv
\frac{\pi\,(2+\partial_y)\,C_s}{2m^2\omega^2\,a_{\ho}^2}\,.
\label{eq:kappa}
\end{equation}

The $Q^{-1}$ scaling follows from the interplay of two exact results: the $q$-independent contact~$C_s$ (from the Laguerre identity) and the $Q$-proportional expectation value of~$R^2$ (from the Laguerre recurrence). If one identifies $C_s$ with the model-operator value, $C_s=\lambda_s/s$, then $\kappa_s$ reduces to a combination of $\lambda_s$, its scale derivative, and the channel parameter~$s$. For the two-body problem, where $\hat{\mathcal C}$ coincides with the physical contact (Sec.~\ref{sec:model}), this identification is exact. For $N>2$, $\kappa_s$ should be treated as a single empirical parameter absorbing contributions not resolved by the single-channel analysis~\cite{WernerCastin2012}.

\begin{figure}[t]
  \centering
  \includegraphics[width=\columnwidth]{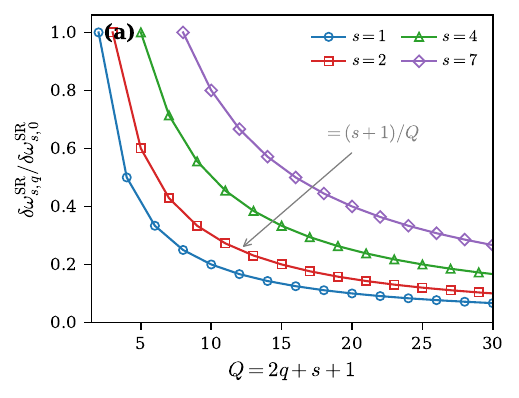}\\[-0.6ex]
  \includegraphics[width=\columnwidth]{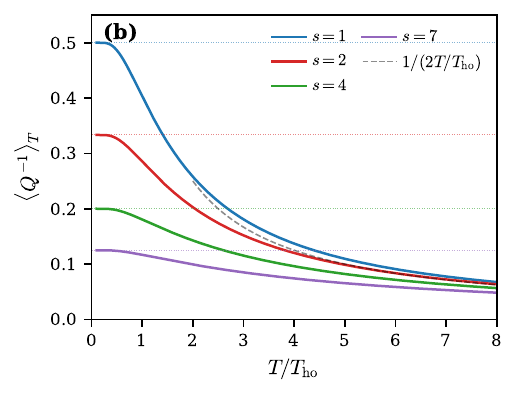}
  \caption{(a) Sum-rule centroid normalized to its value at $q=0$: $\delta\omega^{\rm SR}_{s,q}/\delta\omega^{\rm SR}_{s,0}=(s+1)/Q$, with $Q\equiv 2q+s+1$. This shows the parameter-free $Q^{-1}$ suppression of higher radial levels after a single-point calibration of~$\kappa_s$. (b) Thermal average $\langle Q^{-1}\rangle_T$ weighted by the upward strength $U_{s,q}$, showing the low-$T$ plateau $(s+1)^{-1}$ and the universal high-$T$ tail $1/(2T/T_{\rm ho})$.}
  \label{fig:shift+T}
\end{figure}

At $q=0$ the formula matches the standard sum-rule structure~\cite{Hofmann2012,TaylorRanderia2012,gao2012breathing}; for $q>0$ it resolves the estimate level by level, predicting a monotone suppression with the $Q^{-1}$ scaling (Fig.~\ref{fig:shift+T}a).

\paragraph{Exact line versus sum-rule centroid.} Within the single channel, Eq.~\eqref{eq:exact-2omega} guarantees that the spectral function remains a delta function at $2\omega$ exactly: the single-channel line is \emph{sharp} and \emph{unshifted}. The sum-rule estimate~\eqref{eq:centroid} describes a shifted frequency because the sum-rule bound of Gao and Yu incorporates the scale dependence of the energy (the Tan adiabatic relation~\cite{Tan2008Energetics}) and the equation of state, contributions not visible in the single-channel spectrum. For $N>2$, extracting the observed frequency requires the inter-channel summation discussed in Sec.~\ref{sec:discussion}.

\paragraph{Large-$q$ behavior.} Since $Q=2q+s+1$:
\[
\frac{\delta\omega^{\rm SR}_{s,q}}{2\omega}
=\frac{\kappa_s}{4q}+O(q^{-2}).
\]

\section{First-order spectral weight transfer vanishes}
\label{sec:leakage}

We now present the independent algebraic proof of the first-order spectral-weight cancellation announced in Sec.~\ref{sec:exact-solv}. While the exact solvability of the single-channel model (Sec.~\ref{sec:exact-solv}) already implies that no spectral weight is transferred at any order, the perturbative cancellation mechanism is of independent interest: the ket and bra contributions cancel pairwise, with the $q$- and $s$-dependence dropping out of each pair, revealing algebraic structure that the exact-solvability argument alone does not display. The analogous question in the many-body context---whether the contact perturbation broadens the breathing line---is directly relevant to the interpretation of spectral measurements~\cite{Vogt2012,Holten2018}.

\subsection{Setup}

In the unperturbed system, the operator $R^2$ connects level~$q$ only to $q\pm1$ (tridiagonality). This means the monopole spectral function, viewed from any level~$q$, has all its weight at frequencies $\pm 2\omega$: upward transitions $q\to q{+}1$ at $+2\omega$ and (for $q\ge 1$) downward transitions $q\to q{-}1$ at $-2\omega$. There is no weight at $\pm 4\omega$, $\pm 6\omega$, etc. When the anomaly is turned on, the eigenstates are perturbed: $|\tilde{q}\rangle = |q\rangle + O(\eta)$. The question is whether the perturbed matrix element $\langle\tilde{q}+2|R^2|\tilde{q}\rangle$, which was zero at zeroth order, becomes nonzero at $O(\eta)$.

If it does, there is spectral weight at frequency $\approx 4\omega$ (the energy gap between levels $q$ and $q+2$), and the breathing line would broaden. If it remains zero, the line stays sharp.

\subsection{First-order perturbation theory}

At first order in $\eta$, the perturbed states are
\begin{align}
|\tilde{q}\rangle &= |q\rangle + \eta\sum_{m\neq q}\frac{\langle m|\hat{\mathcal C}|q\rangle}{E_q - E_m}\,|m\rangle,\label{eq:pert-ket}\\
\langle\tilde{q}{+}2| &= \langle q{+}2| + \eta\sum_{n\neq q+2}\frac{\langle q{+}2|\hat{\mathcal C}|n\rangle}{E_{q+2} - E_n}\,\langle n|.\label{eq:pert-bra}
\end{align}
The first-order correction to $\langle q{+}2|R^2|q\rangle$ (which is zero at zeroth order) is
\begin{align}
&\langle\tilde{q}{+}2|R^2|\tilde{q}\rangle^{(1)}
= \eta\bigg(\underbrace{\sum_{m\neq q}\frac{\langle q{+}2|R^2|m\rangle\,\langle m|\hat{\mathcal C}|q\rangle}{E_q - E_m}}_{\text{ket correction}}\notag\\
&\quad+ \underbrace{\sum_{n\neq q+2}\frac{\langle q{+}2|\hat{\mathcal C}|n\rangle\,\langle n|R^2|q\rangle}{E_{q+2} - E_n}}_{\text{bra correction}}\bigg).
\label{eq:first-order}
\end{align}

\subsection{Off-diagonal contact matrix elements}

To evaluate the perturbative sums, we need the off-diagonal elements of $\hat{\mathcal C}$ $\langle q'|\hat{\mathcal C}|q\rangle$ with $q'>q$. These are related to contact matrix elements that also appear in OPE-based treatments of the spectral response~\cite{BraatenPlatter2008OPE}. From the identity~\eqref{eq:Jclosed}, $J^{(s)}_{q',q}$ with $q'>q$ has $\min\{q',q\}=q$, so $J^{(s)}_{q',q} = \Gamma(s)(s{+}1)_q/q!$ (the same value as $J^{(s)}_{q,q}$, independent of how far above $q'$ is). Dividing by the norms gives
\begin{equation}
\langle q{+}\Delta|\hat{\mathcal C}|q\rangle
= \frac{\lambda_s}{s}\,\prod_{j=1}^{\Delta}\sqrt{\frac{q+j}{q+s+j}},
\qquad \Delta = 1,2,3,\ldots
\label{eq:offdiag-contact}
\end{equation}
In particular, the elements needed below are:
\begin{align*}
\langle q{+}1|\hat{\mathcal C}|q\rangle &= \frac{\lambda_s}{s}\sqrt{\frac{q{+}1}{q{+}s{+}1}},\\
\langle q{+}2|\hat{\mathcal C}|q\rangle &= \frac{\lambda_s}{s}\sqrt{\frac{(q{+}1)(q{+}2)}{(q{+}s{+}1)(q{+}s{+}2)}},\\
\langle q{+}3|\hat{\mathcal C}|q\rangle &= \frac{\lambda_s}{s}\sqrt{\frac{(q{+}1)(q{+}2)(q{+}3)}{(q{+}s{+}1)(q{+}s{+}2)(q{+}s{+}3)}}.
\end{align*}
Note the pattern: each factor $\sqrt{(q+j)/(q+s+j)}$ is less than~1 for $s>0$, so the off-diagonal elements decrease with~$|\Delta|$.

\subsection{Identifying the contributing terms}

Because $R^2$ is tridiagonal, $\langle q{+}2|R^2|m\rangle$ is nonzero only for $m \in \{q{+}1, q{+}2, q{+}3\}$. Combined with the constraint $m\neq q$ (which is automatically satisfied for these values), the ket correction has exactly three terms.

Similarly, $\langle n|R^2|q\rangle$ is nonzero only for $n \in \{q{-}1, q, q{+}1\}$, and the constraint $n\neq q{+}2$ is automatically satisfied. The bra correction has at most three terms (two if $q=0$, since $n=q-1=-1$ does not exist).

Inserting the $R^2$ matrix elements from~\eqref{eq:R2matrix}, the contact elements from~\eqref{eq:offdiag-contact}, and the energy denominators $E_q - E_m = -2(m{-}q)\hbar\omega$, each term takes the form $(a_{\ho}^2\lambda_s)/(s\cdot\hbar\omega)$ times a rational function of $q$ and $s$.

\subsection{Worked example: $s=2$, $q=0$}

We demonstrate the cancellation explicitly for the simplest nontrivial case. With $s=2$ and $q=0$ (so $Q=3$, final state $q_f = 2$), the relevant matrix elements are:
\begin{alignat*}{2}
&\langle 1|\hat{\mathcal C}|0\rangle = \frac{\lambda_2}{2\sqrt{3}}, &\qquad
&\langle 2|R^2|1\rangle = -2\sqrt{2}\,a_{\ho}^2,\\
&\langle 2|\hat{\mathcal C}|0\rangle = \frac{\lambda_2}{2\sqrt{6}}, &\qquad
&\langle 2|R^2|2\rangle = 7\,a_{\ho}^2,\\
&\langle 3|\hat{\mathcal C}|0\rangle = \frac{\lambda_2}{2\sqrt{10}}, &\qquad
&\langle 2|R^2|3\rangle = -\sqrt{15}\,a_{\ho}^2,\\
&\langle 2|\hat{\mathcal C}|1\rangle = \frac{\lambda_2}{2\sqrt{2}}, &\qquad
&\langle 0|R^2|0\rangle = 3\,a_{\ho}^2,\\
& &\qquad
&\langle 1|R^2|0\rangle = -\sqrt{3}\,a_{\ho}^2.
\end{alignat*}
With the common prefactor $\mathcal{P} \equiv \lambda_2 a_{\ho}^2/(4\sqrt{6}\,\hbar\omega)$ (note that $\eta$ is already factored out in Eq.~\eqref{eq:first-order}), the five terms contributing to the bracket are:
\begin{alignat*}{2}
&\text{Ket } m{=}1\!: &\quad &+\,4\,\mathcal{P},\\
&\text{Ket } m{=}2\!: &\quad &-\,\tfrac{7}{2}\,\mathcal{P},\\
&\text{Ket } m{=}3\!: &\quad &+\,1\,\mathcal{P},\\
&\text{Bra } n{=}0\!: &\quad &+\,\tfrac{3}{2}\,\mathcal{P},\\
&\text{Bra } n{=}1\!: &\quad &-\,3\,\mathcal{P}.
\end{alignat*}
The ket sum is $(4 - 7/2 + 1)\,\mathcal{P} = +3/2\,\mathcal{P}$ and the bra sum is $(3/2 - 3)\,\mathcal{P} = -3/2\,\mathcal{P}$, giving zero exactly. The term $n=q{-}1=-1$ is absent at $q=0$; for $q\ge 1$ there is a sixth term, and the cancellation still holds (verified for all $s$ and $q$ tested, see Sec.~\ref{sec:validation}).

\subsection{General cancellation}

The cancellation demonstrated above for $s=2$, $q=0$ persists for all $s>0$ and $q\ge 0$:
\begin{equation}
\langle\tilde{q}{+}2|R^2|\tilde{q}\rangle^{(1)} = 0\qquad\text{for all }s>0,\;q\ge0.
\label{eq:zero-leak}
\end{equation}
The same argument applies to $\langle\tilde{q}{-}2|R^2|\tilde{q}\rangle$ (the downward $q\to q{-}2$ channel). \emph{There is no transfer of spectral weight at first order in the anomaly.}

In fact, the general first-order amplitude can be summed in closed form. Factoring out
\[
\mathcal A_{s,q}\equiv
\frac{\eta\lambda_s a_{\ho}^2}{s\hbar\omega}
\sqrt{\frac{(q{+}1)(q{+}2)}{(q{+}s{+}1)(q{+}s{+}2)}}\,,
\]
the ket correction equals
\begin{align*}
&\mathcal A_{s,q}\!\left[
\frac{q{+}s{+}2}{2}
-\frac{2q{+}s{+}5}{4}
+\frac{q{+}3}{6}
\right]\\
&\qquad=
\mathcal A_{s,q}\,\frac{2q+3s+3}{12}\,,
\end{align*}
while the bra correction equals
\begin{align*}
&\mathcal A_{s,q}\!\left[
-\frac{q}{6}
+\frac{2q{+}s{+}1}{4}
-\frac{q{+}s{+}1}{2}
\right]\\
&\qquad=
-\mathcal A_{s,q}\,\frac{2q+3s+3}{12}\,.
\end{align*}
(The $q=0$ case is included automatically, since the $-q/6$ term vanishes.) The two pieces cancel exactly, making the result~\eqref{eq:zero-leak} fully analytic.

\paragraph{Extension to all $\ell\ge 2$.} The same cancellation holds for every forbidden channel $q\to q+\ell$ with $\ell\ge 2$. We give the explicit derivation.

Because $R^2$ is tridiagonal, the first-order amplitude~\eqref{eq:first-order} receives three ket terms ($m=q{+}\ell{-}1,\,q{+}\ell,\,q{+}\ell{+}1$) and three bra terms ($n=q{-}1,\,q,\,q{+}1$). For $\ell\ge 2$ no exclusions $m\neq q$ or $n\neq q{+}\ell$ are needed. Factor out $\mathcal P \equiv \eta\,c_\ell(q)\,a_{\ho}^2/(2\hbar\omega)$ with $c_\ell(q)\equiv\langle q{+}\ell|\hat{\mathcal C}|q\rangle$.

\emph{Ket terms.} The energy denominators are $E_q-E_m = -2(m{-}q)\hbar\omega$. Each term involves a product of an $R^2$ element and a contact ratio $c_\delta(q)/c_\ell(q)$. Using Eq.~\eqref{eq:offdiag-contact}, the contact ratios simplify via $c_{\ell-1}(q)/c_\ell(q) = \sqrt{(q{+}s{+}\ell)/(q{+}\ell)}$ and $c_{\ell+1}(q)/c_\ell(q)=\sqrt{(q{+}\ell{+}1)/(q{+}s{+}\ell{+}1)}$, so the products of $R^2$ elements and contact ratios collapse (using $\sqrt{a/b}\cdot\sqrt{ab}=a$):
\begin{alignat*}{2}
&\text{(K1)}\;m{=}q{+}\ell{-}1: \quad &\tfrac{-(q{+}s{+}\ell)}{-2(\ell{-}1)} &= \tfrac{q{+}s{+}\ell}{2(\ell{-}1)},\\
&\text{(K2)}\;m{=}q{+}\ell: &\tfrac{Q{+}2\ell}{-2\ell} &= -\tfrac{Q{+}2\ell}{2\ell},\\
&\text{(K3)}\;m{=}q{+}\ell{+}1: &\tfrac{-(q{+}\ell{+}1)}{-2(\ell{+}1)} &= \tfrac{q{+}\ell{+}1}{2(\ell{+}1)}.
\end{alignat*}

\emph{Bra terms.} The energy denominators are $E_{q+\ell}-E_n = 2(q{+}\ell{-}n)\hbar\omega$. For the contact ratios with shifted base point, the same product pattern gives $c_{\ell+1}(q{-}1)/c_\ell(q) = \sqrt{q/(q{+}s)}$ and $c_{\ell-1}(q{+}1)/c_\ell(q) = \sqrt{(q{+}s{+}1)/(q{+}1)}$:
\begin{alignat*}{2}
&\text{(B1)}\;n{=}q{-}1: \quad &\tfrac{-q}{2(\ell{+}1)} &= -\tfrac{q}{2(\ell{+}1)},\\
&\text{(B2)}\;n{=}q: &\tfrac{Q}{2\ell} &= \tfrac{Q}{2\ell},\\
&\text{(B3)}\;n{=}q{+}1: &\tfrac{-(q{+}s{+}1)}{2(\ell{-}1)} &= -\tfrac{q{+}s{+}1}{2(\ell{-}1)}.
\end{alignat*}

The six terms pair naturally by energy denominator:
\begin{align*}
\text{pair }(\ell{-}1)&:\;\text{K1}{+}\text{B3} = \tfrac{(q{+}s{+}\ell)-(q{+}s{+}1)}{2(\ell{-}1)}=\tfrac{1}{2},\\[3pt]
\text{pair }(\ell)\phantom{+1}&:\;\text{K2}{+}\text{B2} = \tfrac{-(Q{+}2\ell)+Q}{2\ell}=-1,\\[3pt]
\text{pair }(\ell{+}1)&:\;\text{K3}{+}\text{B1} = \tfrac{(q{+}\ell{+}1)-q}{2(\ell{+}1)}=\tfrac{1}{2}.
\end{align*}
Each pair collapses to a constant independent of $q$ and~$s$, and the total is $\frac{1}{2}-1+\frac{1}{2}=0$. (At $q=0$ the $n=q{-}1$ term is absent, but so is the $-q/(2(\ell{+}1))$ contribution.) The downward channels $q\to q{-}\ell$ follow by the same argument. Thus
\begin{equation}
\langle\tilde{q}{+}\ell|R^2|\tilde{q}\rangle^{(1)} = 0\qquad\text{for all }\ell\ge 2,\;s>0,\;q\ge0,
\label{eq:zero-leak-general}
\end{equation}
which is the full statement of selection-rule preservation: the contact perturbation transfers no monopole spectral weight away from $\pm2\omega$ at first order.

\emph{Remark on normalization.} The first-order kets used in~\eqref{eq:first-order} are not normalized. However, since $\langle\tilde{q}|\tilde{q}\rangle = 1 + O(\eta^2)$, the normalization correction enters only at $O(\eta^2)$; moreover, it multiplies the zeroth-order matrix element $\langle q{+}\ell|R^2|q\rangle = 0$, so it cannot generate spectral weight at forbidden frequencies.

\subsection{Physical interpretation}

Why is spectral weight not redistributed? One ingredient is the $q$-independence of the diagonal contact, Eq.~\eqref{eq:contact-diag}. Because $\langle\hat{\mathcal C}\rangle_{s,q} = \lambda_s/s$ for all~$q$, the contact perturbation shifts all levels in a channel by the same amount. A uniform level shift removes any first-order gap-shift mechanism within the channel.

The residue correction, however, is controlled by the \emph{off-diagonal} contact elements $\langle q'|\hat{\mathcal C}|q\rangle$ with $q'\neq q$. Their special structure (inherited from the gamma-ratio identity~\eqref{eq:Jclosed}) conspires with the tridiagonality of~$R^2$ to produce exact cancellation between ket and bra corrections. The vanishing spectral weight transfer is therefore a consequence of both the uniform diagonal shift and the closed off-diagonal pattern.

The physical consequence is that the breathing line remains spectrally sharp: the exact solvability (Sec.~\ref{sec:exact-solv}) guarantees this at all orders, while the perturbative cancellation proved here confirms the mechanism at $O(\eta)$. The sum-rule centroid~\eqref{eq:centroid} describes a shifted frequency arising from the many-body equation of state, but the single-channel line itself does not broaden or shift. This is consistent with both the theoretical expectation~\cite{Chiacchiera2013_PRA_quadDamping2D,Son2007PRL_BulkViscosity} and the experimental observation that the breathing mode remains a well-defined, narrow spectral feature even in the presence of measurable anomalous shifts~\cite{Vogt2012,Holten2018,Peppler2018,Murthy2019}.

\subsection{Numerical validation}
\label{sec:validation}

\paragraph{First-order perturbation theory.} We evaluate the ket and bra sums in~\eqref{eq:first-order} numerically using the analytic matrix elements of $\hat{\mathcal C}$ from~\eqref{eq:Jclosed}. For all tested values of $s$ and $q$ ($s=1,2,4$; $q=0,1,2,5,10$), the ket and bra sums cancel to machine precision, with residuals $\lesssim 10^{-15}$.

\begin{figure}[t]
  \centering
  \includegraphics[width=\columnwidth]{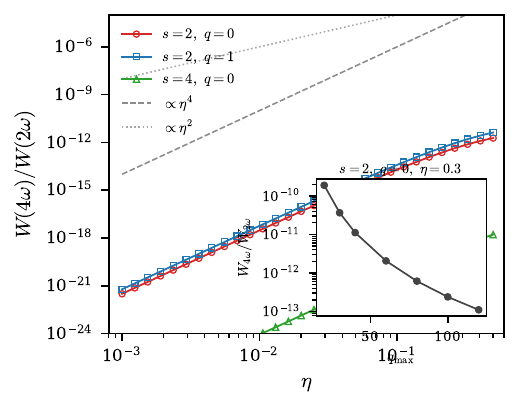}
  \caption{Spectral weight at $4\omega$ relative to the weight at $2\omega$, from exact diagonalization. The $\eta^4$ scaling (dashed line) confirms that the first-order transfer amplitude vanishes; the dotted line shows $\eta^2$, the scaling that would obtain if spectral weight were redistributed at first order. Inset: basis-size dependence of the residual $4\omega$ weight for $s=2$, $q=0$, $\eta=0.3$. The rapid decrease with $q_{\max}$ confirms that the nonzero value in the truncated unperturbed basis is a numerical artifact; in the exact single-channel model with shifted parameter $s_\eta$ the forbidden weight vanishes identically (Sec.~\ref{sec:exact-solv}).}
  \label{fig:leakage}
\end{figure}

\paragraph{Exact diagonalization.}
We construct the full Hamiltonian $H = H_0 + \eta\hat{\mathcal C}$ as a matrix in the truncated basis $\{|s,q\rangle : q = 0,\ldots,q_{\max}\}$ with $q_{\max}=80$, diagonalize it, and compute the $R^2$ spectral function from each eigenstate. Figure~\ref{fig:leakage} shows the spectral weight at frequency~$4\omega$ (the $q\to q{+}2$ channel), normalized by the weight at~$2\omega$, as a function of~$\eta$ for several values of $(s,q)$.

The $\eta^4$ scaling at small $\eta$ is the key signature: since the spectral weight is $|\langle\tilde{q}+2|R^2|\tilde{q}\rangle|^2$, an $\eta^4$ scaling means the amplitude scales as $\eta^2$, i.e., the first-order ($O(\eta)$) amplitude vanishes and the leading contribution is second-order. (In the exact single-channel model with inverse-square perturbation of Sec.~\ref{sec:exact-solv}, the forbidden amplitude vanishes at \emph{all} orders; the residual $\eta^4$ weight visible in Fig.~\ref{fig:leakage} arises from working in the \emph{unperturbed} basis rather than the exact $s_\eta$ basis.) The deviations from $\eta^4$ at larger~$\eta$ are expected from higher-order terms in the unperturbed-basis expansion. This provides independent confirmation of the analytic result~\eqref{eq:zero-leak}.

\section{Finite-temperature sum-rule estimate}
\label{sec:finiteT}

At finite temperature, multiple radial levels within a channel are thermally populated. The sum-rule centroid~\eqref{eq:centroid} assigns each level $q$ a shift $\delta\omega_{\rm cen}^{(q)}/(2\omega) = \kappa_s/(2Q)$. We stress that Eq.~\eqref{eq:centroid} is not the exact spectral line position of the single-channel Hamiltonian $H_{s,\eta}$ (whose spectrum stays at $\pm 2\omega$ exactly, Eq.~\eqref{eq:exact-2omega}), but the sum-rule centroid for the channel derived in Sec.~\ref{sec:centroid-derivation}. A natural thermal average of these per-level values is obtained by weighting each level by its Boltzmann factor $p_q \propto x^q$ and its upward oscillator strength $U_{s,q}$~\cite{LippariniStringari1989,Stringari1996}. (In the retarded susceptibility $\chi''(\omega)$, detailed balance introduces the factor $(p_q - p_{q+1}) = (1-x)\,p_q$; because all lines sit at the same frequency~$2\omega$, this common factor $(1-x)$ cancels in a normalized centroid.) The resulting thermal centroid within channel~$s$ is
\begin{equation}
\frac{\delta\omega^{(s)}_{\rm cen}(T)}{2\omega}
=\frac{\kappa_s}{2}\,\big\langle Q^{-1}\big\rangle_T,
\label{eq:thermalMain}
\end{equation}
where
\begin{equation}
\big\langle Q^{-1}\big\rangle_T \equiv \frac{\sum_{q\ge0}U_{s,q}\,Q^{-1}\,x^q}{\sum_{q\ge0}U_{s,q}\,x^q},
\qquad x\equiv e^{-2\hbar\omega/(k_BT)}.
\label{eq:thermal-avg}
\end{equation}
The weight $U_{s,q}$ is the correct choice for a positive-frequency centroid of the positive-frequency spectrum: each physical transition $q\to q{+}1$ is counted once, weighted by its upward oscillator strength from the occupied level~$q$. (Using the total unsigned strength $W_{s,q} = U_{s,q}+D_{s,q}$ would double-count: the same transition $q\leftrightarrow q{+}1$ would appear once through $U_{s,q}$ in level~$q$'s weight and again through $D_{s,q+1}=U_{s,q}$ in level~$q{+}1$'s weight, but assigned different centroid factors $Q_q^{-1}$ and $Q_{q+1}^{-1}$.)

This is a thermal average \emph{within a single channel}; the factor $x^q = e^{-\beta(E_{s,q}-E_{s,0})}$ is the canonical Boltzmann weight within the radial ladder and does not presume noninteracting single-particle statistics. The connection to the many-body breathing response requires summation over channels weighted by their many-body occupation (see Sec.~\ref{sec:discussion}).

The denominator sums to a closed rational function:
\begin{equation}
\sum_{q\ge0} U_{s,q}\,x^q = \frac{(s{+}1)+(1{-}s)\,x}{(1-x)^3}.
\label{eq:Usx}
\end{equation}
The numerator admits a compact representation via the Lerch transcendent $\Phi(x,\nu,a)\equiv\sum_{k\ge0}x^k/(k+a)^\nu$~\cite{Lewin1981,HTF1}:
\[
\sum_{q\ge0}\frac{U_{s,q}}{Q}\,x^q
= \frac{x}{2(1{-}x)^2}+\frac{s{+}3}{4(1{-}x)}
-\frac{s^2{-}1}{8}\,\Phi\!\left(x,1,\tfrac{s{+}1}{2}\right).
\]
The limits of the thermal average are:
\[
\big\langle Q^{-1}\big\rangle_T \;\to\;
\begin{cases}
(s{+}1)^{-1} & T\to 0,\\[0.5ex]
(1{-}x)/4 \simeq \hbar\omega/(2k_BT) & T\to\infty.
\end{cases}
\]
At low temperature, only $q=0$ is occupied, and $U_{s,0}/(2\cdot0+s+1) = (s{+}1)/(s{+}1) = 1$, giving $\langle Q^{-1}\rangle_T = 1/(s{+}1)$. At high temperature, the dominant levels have $q\sim k_BT/(2\hbar\omega)$, so $Q^{-1}\sim \hbar\omega/(k_BT)$ and the estimate decays as~$1/T$. Physically, heating populates higher radial levels where the centroid contribution is suppressed by the $Q^{-1}$ factor (Fig.~\ref{fig:shift+T}b).

\section{Robustness under weak anisotropy}
\label{sec:anisotropy}

Real traps are never perfectly isotropic. We now discuss the extent to which the key results survive under weak anisotropy.

Let the squared trap frequencies be $\omega_i^2=\omega^2(1+\epsilon_i)$ with $\sum_i\epsilon_i=0$ and $|\epsilon_i|\ll1$. The anisotropic perturbation can be written as
\[
\delta V=\tfrac12 m\omega^2\,\epsilon_2\,R^2\,\hat{\mathcal Q},
\]
where $\epsilon_2 \equiv (\sum_i \epsilon_i^2)^{1/2}$ is the overall anisotropy magnitude and $\hat{\mathcal Q}(\Omega) \equiv \epsilon_2^{-1}\sum_i \epsilon_i\,r_i^2/R^2$ is a traceless angular operator. The key structural point is that $\delta V$ factors into a radial piece ($R^2$) times an angular piece ($\hat{\mathcal Q}$). Consequently, the radial matrix elements of $\delta V$ have exactly the same three-term tridiagonal structure as $R^2$ itself:
Within a single channel ($s'=s$), the radial part of $\delta V$ inherits the tridiagonality of~$R^2$:
\[
\langle s,q'|\,\delta V\,|s,q\rangle
=\tfrac12 m\omega^2 a_{\ho}^2\,\epsilon_2\,\langle s|\hat{\mathcal Q}|s\rangle
\,\langle q'|R^2/a_{\ho}^2|q\rangle_{s},
\]
where $\langle q'|R^2/a_{\ho}^2|q\rangle_{s}$ is the fixed-$s$ tridiagonal matrix from Eq.~\eqref{eq:R2matrix}. For cross-channel matrix elements ($s'\neq s$), the radial wavefunctions depend on different values of the channel parameter, and the $R^2$ matrix element $\langle q'(s')|R^2|q(s)\rangle$ is not in general tridiagonal. We do not evaluate these cross-channel radial overlaps here.

This has two consequences and one important limitation.

First, \emph{within a single channel}, the anisotropy perturbation does not break the radial selection rule $\Delta q = \pm 1$, because the radial part of $\delta V$ inherits the tridiagonality of~$R^2$.

Second, the anomaly perturbation \emph{by itself} preserves $\Delta q = \pm 1$ (Eq.~\eqref{eq:zero-leak-general}), as proved in Sec.~\ref{sec:leakage}.

The limitation is that the anisotropy mixes different angular channels through $\langle s'|\hat{\mathcal Q}|s\rangle \neq 0$ for $s'\neq s$. Once channels are mixed, the eigenstates are no longer pure $|s,q\rangle$ states, and the notion of ``within a single channel'' becomes approximate. Our zero-leakage proof relied on the specific matrix elements of $\hat{\mathcal C}$ \emph{within} a single channel; for the combined anomaly-plus-anisotropy perturbation, the intermediate states in the perturbative sums include other-channel contributions whose contact structure is different. We have not proved that the cancellation persists for the combined perturbation, and at $O(\varepsilon_A\,\epsilon_2)$ mixed terms may appear.

For weak anisotropy ($|\epsilon_2| \ll 1$), the channel mixing is itself a small correction, so the within-channel results are expected to hold approximately. A quantitative analysis of the combined anomaly--anisotropy case is left for future work~\cite{GueryOdelin1999PRA,Stringari1996,KaganSurkovShlyapnikov1996PRA,KaganSurkovShlyapnikov1997PRA,Lobser2015}.

\section{Calibration and experimental connection}
\label{sec:calibration}

\subsection{Calibration procedure}

The exact single-channel result of this paper is Eq.~\eqref{eq:exact-2omega}: the monopole line remains at $2\omega$ exactly within the single-channel model. The sum-rule estimate~\eqref{eq:centroid} is derived by substituting single-channel quantities into the sum-rule bound of Gao and Yu (Sec.~\ref{sec:centroid-derivation}); the coefficient $\kappa_s$ defined in~\eqref{eq:kappa} involves the contact $C_s$ and its scale derivative, which are not computed from the single-channel model alone.

A practical route is to calibrate $\kappa_s$ from a low-temperature or $q{=}0$ measurement of the breathing frequency. For the two-body problem, the identification $C_s=\lambda_s/s$ is exact; for many-body experiments where a single channel dominates, the fitted $\kappa_s$ absorbs the (unknown) channel-weighting structure, finite-range effects, and other contributions. With $\kappa_s$ calibrated from a single point, the testable predictions are the \emph{functional form} of the $q$-dependence ($1/Q$ scaling) and the $T$-dependence (the shape of the $\langle Q^{-1}\rangle_T$ curve, including the crossover from the low-$T$ plateau to the $1/T$ tail). Agreement or disagreement with these functional forms tests the single-channel sum-rule framework independently of the overall magnitude.

Note that a single measurement at $q{=}0$ fixes $\kappa_s$ directly; separating the contact and curvature contributions requires either an independent estimate or a second measurement in a different channel.

\subsection{Worked example ($s=2$)}

For the commonly used $s=2$ channel at $q{=}0$ ($Q=3$), Eq.~\eqref{eq:centroid} gives
\[
\frac{\delta\omega^{\rm SR}_{2,0}}{2\omega}
=\frac{\kappa_2}{6}\,.
\]
A single reference measurement at $q{=}0$ therefore fixes $\kappa_2 = 6\,[\delta\omega^{\rm SR}/(2\omega)]_{q=0,s=2}$. The estimate for any $q$ in the same channel is then
\[
\frac{\delta\omega^{\rm SR}_{2,q}}{2\omega}=\frac{\kappa_2}{2Q}\,,\qquad Q=2q+3,
\]
and the thermal average follows from~\eqref{eq:thermalMain} with no additional free parameters.

Representative 2D $^6$Li measurements report fractional shifts $|\delta\omega|/(2\omega)\!\sim\!1\%$--$5\%$~\cite{Peppler2018,Holten2018,Murthy2019}. For $s{=}2$, $q{=}0$, this corresponds to $|\kappa_2|\!\approx\!0.06$--$0.30$.

\subsection{Finite range and quasi-2D}

Quasi-2D realizations induce an effective range that weakens the anomaly relative to ideal 2D~\cite{hu2019reduced,Yin2020_PRL_fewbodyAnomaly2D}. In the sum-rule centroid for the channel~\eqref{eq:centroid}, these effects enter through the contact $C_s$ and its scale derivative, which acquire geometry dependence. This is absorbed into a geometry-dependent coefficient, $\kappa_s\!\to\!\kappa_s(r_{\rm eff},\ell_z)$, so that
\[
\frac{\delta\omega^{\rm SR}_{s,q}}{2\omega}=\frac{\kappa_s(r_{\rm eff},\ell_z)}{2Q}\,.
\]
The $Q^{-1}$ structure is preserved because it originates from the $Q$-proportional expectation value of $R^2$, which is a single-channel kinematic result independent of the interaction details. A calibration at $q{=}0$ in the same geometry then fixes the remaining $q$- and $T$-dependence within that channel~\cite{hu2019reduced,Yin2020_PRL_fewbodyAnomaly2D}.

\section{Discussion and outlook}
\label{sec:discussion}

\subsection{Summary of results}

We have shown that the $1/R^2$ perturbation within a single hyperangular channel of a harmonically trapped system is absorbed exactly into a shift of the channel parameter, $s\to s_\eta$ (Eq.~\eqref{eq:seta}), so the single-channel model remains a harmonic oscillator with a shifted inverse-square term at all orders. A classical identity for products of associated Laguerre polynomials~\cite{Gasper1975,SrivastavaMavromatisAlassar2003} provides closed-form matrix elements. The exact consequences are:

\begin{enumerate}
\item \emph{Exact $2\omega$ spacing} (Eq.~\eqref{eq:exact-2omega}): radial gaps remain $2\hbar\omega$ exactly, $R^2$ stays tridiagonal in the exact eigenbasis, and no monopole spectral weight appears at forbidden frequencies at any order in~$\eta$.
\item \emph{$q$-independent diagonal matrix element} (Eq.~\eqref{eq:contact-diag}): $\langle\hat{\mathcal C}\rangle_{s,q}=\lambda_s/s$ for all~$q$, where $\hat{\mathcal C}$ is the $1/R^2$-weighted single-channel operator defined in Sec.~\ref{sec:model}.
\item \emph{Independent first-order algebraic proof} (Eqs.~\eqref{eq:zero-leak}--\eqref{eq:zero-leak-general}): the first-order cancellation of forbidden spectral weight is proved by an explicit term-by-term cancellation in which the $q$- and $s$-dependence drops out of each pair, confirmed by exact diagonalization (Fig.~\ref{fig:leakage}).
\end{enumerate}

Substituting the exact single-channel quantities into the $m_1/m_{-1}$ sum-rule bound of Gao and Yu~\cite{gao2012breathing} yields a sum-rule estimate with $Q^{-1}$ scaling (Eq.~\eqref{eq:centroid}) and an explicit coefficient~\eqref{eq:kappa}. Its finite-temperature average (Eq.~\eqref{eq:thermalMain}), using upward oscillator strengths $U_{s,q}$ as the correct positive-frequency weights~\cite{LippariniStringari1989}, has a low-$T$ plateau and $1/T$ high-$T$ tail with closed-form expressions involving the Lerch transcendent~\cite{Lewin1981}.

The Laguerre identity, the ratio $I_{s-1}/I_s=1/s$, the exact solvability, and the selection-rule preservation hold for any real $s>0$ and thus apply formally in three dimensions upon $s\to\alpha$; the sum-rule-estimate interpretation, however, encounters an additional complication in 3D (see below).

\subsection{Single-channel results and the many-body breathing mode}

Several layers of interpretation separate the exact single-channel results proved here from a measured many-body breathing frequency.

First, the exact single-channel statement is Eq.~\eqref{eq:exact-2omega}: the radial spectrum has exact $2\omega$ spacing at all orders in~$\eta$, and no forbidden spectral weight appears. This is a rigorous consequence of the inverse-square structure of the model Hamiltonian~\eqref{eq:Hseta}.

Second, many-body sum-rule and hydrodynamic approaches~\cite{Hofmann2012,TaylorRanderia2012,gao2012breathing} relate the breathing-mode shift to the equation of state and contact derivatives. The single-channel sum-rule estimate~\eqref{eq:centroid} is derived by inserting exact single-channel quantities ($q$-independent contact $C_s$, $Q$-proportional $\langle R^2\rangle$) into the sum-rule bound of Gao and Yu. The $Q^{-1}$ structure follows from these two inputs; the overall coefficient~$\kappa_s$ depends on the contact $C_s$ and its scale derivative, which for $N>2$ are not computed from the single-channel model alone.

Third, the finite-$T$ average~\eqref{eq:thermalMain} is a thermal average of the derived per-level sum-rule estimates within one channel. The actual many-body finite-$T$ response depends on the full density matrix across all channels and on inter-channel couplings. The single-channel formula captures the qualitative behavior (low-$T$ plateau, $1/T$ high-$T$ tail), but its quantitative comparison with experiment requires the inter-channel summation that we do not perform.

The main open problem is therefore the bridge from exact single-channel matrix elements to the full many-body response. For the two-body problem this bridge is trivial for the exact results (matrix elements, selection rule, exact solvability); the sum-rule centroid can be compared with the exact two-body frequency from Busch's solution~\cite{Busch1998}. For $N>2$ the bridge requires knowledge of the many-body channel occupation, which depends on the interaction strength and statistics.

\subsection{Outlook}

The most direct tests of the exact results derived here are few-body and state-resolved. In optical microtraps or tweezer platforms, where individual motional levels of two atoms can be prepared and probed~\cite{Serwane2011Science,Zurn2012PRL,Hood2020PRResearch}, one may test the closed-form matrix elements, the absence of forbidden monopole transitions (exact at all orders within the single-channel model), and the sum-rule centroid~\eqref{eq:centroid} against the exact two-body frequency~\cite{Busch1998}.

For many-body experiments, the sum-rule centroid~\eqref{eq:centroid} provides a compact one-parameter scheme once a reference point is supplied. In trap units, the high-$T$ slope of $\delta\omega_{\rm cen}^{(s)}(T)/(2\omega)$ versus $1/(T/T_{\rm ho})$ is $\kappa_s/4$, while the low-$T$ plateau tends to $\kappa_s/(2(s+1))$. These functional forms are parameter-free once $\kappa_s$ is calibrated at $T\!\approx\!0$; their agreement or disagreement with a temperature scan of $\omega_B(T)$---in the spirit of recent precision studies of 2D collective dynamics~\cite{He2020,Bohlen2020}---would provide a direct test of the single-channel sum-rule framework. Because exact scale invariance implies vanishing bulk viscosity, any dissipative broadening remains a sensitive symmetry-breaking diagnostic~\cite{Son2007PRL_BulkViscosity,Enss2019PRL_BulkViscosityContact}.

The Laguerre polynomial identities and the exact solvability extend formally to any real positive parameter, so the integral identities, the exact $2\omega$ radial spacing, and the cancellation mechanism apply formally in three dimensions upon $s\to\alpha$, where $\alpha$ is the 3D hyperradial parameter~\cite{WernerCastin2006,CastinNotes2011}: the identity $I_{\alpha-1}/I_\alpha=1/\alpha$, the off-diagonal matrix element formula~\eqref{eq:offdiag-contact}, the selection-rule preservation~\eqref{eq:zero-leak-general}, and the exact spectrum~\eqref{eq:exact-spectrum} all hold with $s$ replaced by~$\alpha$. We have verified the selection-rule preservation and the ratio $I_{\alpha-1}/I_\alpha=1/\alpha$ numerically for non-integer values $\alpha = 0.5, 1.5, 2.5, 3.7$.

What does \emph{not} follow straightforwardly is the physical interpretation in three dimensions. Werner and Castin~\cite{WernerCastin2012} showed that within an $SO(2,1)$ energy ladder of a trapped 3D gas, the energy derivatives $\partial E/\partial(1/a)$ and $\partial E/\partial r_e$ vary with the ladder index~$q$. This $q$-dependence of the physical contact correction is incompatible with the $q$-independent model operator $\langle\hat{\mathcal C}\rangle_{s,q}=\lambda_s/s$. A genuine three-dimensional breathing-mode application therefore requires a separate derivation that incorporates the $q$-dependent corrections. At exact 3D unitarity the anomaly piece vanishes ($\varepsilon_A=0$)~\cite{WernerCastin2006,TaylorRanderia2012}, and the exact results remain valid and potentially useful as building blocks for a more complete treatment.

In the presence of weak quartic anharmonicity ($V_{\rm ext}=\tfrac12 m\omega^2 r^2 + \kappa r^4$), the anharmonicity breaks the uniform level spacing and can shift the single-channel line independently of the inverse-square perturbation studied here~\cite{DeSilva2008PRA}; combining this effect with the anomaly in the many-body sum-rule framework yields a mixed term~\cite{TaylorRanderia2012,Stringari1996} whose coefficient we have not independently verified within the single-channel model.

\bibliographystyle{apsrev4-2-titles}
\bibliography{ufg_refs}

@article{WernerCastin2006,
  title   = {Unitary gas in an isotropic harmonic trap: {S}ymmetry properties and applications},
  author  = {Werner, F{\'e}lix and Castin, Yvan},
  journal = {Phys. Rev. A},
  volume  = {74},
  pages   = {053604},
  year    = {2006},
  doi     = {10.1103/PhysRevA.74.053604},
  url     = {https://link.aps.org/doi/10.1103/PhysRevA.74.053604}
}

@article{CastinNotes2011,
  title   = {The Unitary Gas and its Symmetry Properties},
  author  = {Castin, Yvan},
  journal = {arXiv:1103.2851},
  year    = {2011},
  url     = {https://arxiv.org/abs/1103.2851}
}

@article{WernerCastin2012,
  title   = {General relations for quantum gases in two and three dimensions: Two-component fermions},
  author  = {Werner, F{\'e}lix and Castin, Yvan},
  journal = {Phys. Rev. A},
  volume  = {86},
  pages   = {013626},
  year    = {2012},
  doi     = {10.1103/PhysRevA.86.013626},
  url     = {https://link.aps.org/doi/10.1103/PhysRevA.86.013626}
}

@article{PitaevskiiRosch1997,
  title   = {Breathing modes and hidden symmetry of trapped atoms in two dimensions},
  author  = {Pitaevskii, Lev P. and Rosch, Achim},
  journal = {Phys. Rev. A},
  volume  = {55},
  number  = {R},
  pages   = {R853--R856},
  year    = {1997},
  doi     = {10.1103/PhysRevA.55.R853},
  url     = {https://link.aps.org/doi/10.1103/PhysRevA.55.R853}
}

@article{SrivastavaMavromatisAlassar2003,
  title   = {Remarks on Some Associated {L}aguerre Integral Results},
  author  = {Srivastava, Hari Mohan and Mavromatis, Harry A. and Alassar, Raid S.},
  journal = {Applied Mathematics Letters},
  volume  = {16},
  number  = {7},
  pages   = {1131--1136},
  year    = {2003},
  doi     = {10.1016/S0893-9659(03)90106-6}
}

@article{Busch1998,
  title   = {Two Cold Atoms in a Harmonic Trap},
  author  = {Busch, Thomas and Englert, Berthold-Georg and Rza{\.z}ewski, Kazimierz and Wilkens, Martin},
  journal = {Foundations of Physics},
  volume  = {28},
  number  = {4},
  pages   = {549--559},
  year    = {1998},
  doi     = {10.1023/A:1018705520999}
}

@article{TaylorRanderia2012,
  title   = {Apparent Low-Energy Scale Invariance in two-dimensional {F}ermi gases},
  author  = {Taylor, Edward and Randeria, Mohit},
  journal = {Phys. Rev. Lett.},
  volume  = {109},
  pages   = {135301},
  year    = {2012},
  doi     = {10.1103/PhysRevLett.109.135301},
  url     = {https://link.aps.org/doi/10.1103/PhysRevLett.109.135301}
}

@article{TaylorRanderia2013Erratum,
  title   = {Erratum: Apparent Low-Energy Scale Invariance in Two-Dimensional {F}ermi Gases},
  author  = {Taylor, Edward and Randeria, Mohit},
  journal = {Phys. Rev. Lett.},
  volume  = {110},
  pages   = {089904},
  year    = {2013},
  doi     = {10.1103/PhysRevLett.110.089904},
  url     = {https://link.aps.org/doi/10.1103/PhysRevLett.110.089904}
}

@article{Hofmann2012,
  title   = {Quantum Anomaly, Universal Relations, and Breathing Mode of a Two-Dimensional {F}ermi Gas},
  author  = {Hofmann, Johannes},
  journal = {Phys. Rev. Lett.},
  volume  = {108},
  pages   = {185303},
  year    = {2012},
  doi     = {10.1103/PhysRevLett.108.185303},
  url     = {https://link.aps.org/doi/10.1103/PhysRevLett.108.185303}
}

@article{Castin2004,
  author  = {Castin, Yvan},
  title   = {Exact scaling transform for a unitary quantum gas in a time dependent harmonic potential},
  journal = {C. R. Physique},
  volume  = {5},
  pages   = {407--410},
  year    = {2004},
  doi     = {10.1016/j.crhy.2004.03.017}
}

@article{Vogt2012,
  author  = {Vogt, Emanuel and Feld, Michael and Fr{\"o}hlich, Benjamin and Pertot, Daniel and Koschorreck, Marco and K{\"o}hl, Michael},
  title   = {Scale Invariance and Viscosity of a Two-Dimensional {F}ermi Gas},
  journal = {Phys. Rev. Lett.},
  volume  = {108},
  pages   = {070404},
  year    = {2012},
  doi     = {10.1103/PhysRevLett.108.070404}
}

@article{Holten2018,
  author  = {Holten, Marvin and Bayha, Luca and Klein, Antonia C. and Murthy, Puneet A. and Preiss, Philipp M. and Jochim, Selim},
  title   = {Anomalous Breaking of Scale Invariance in a Two-Dimensional {F}ermi Gas},
  journal = {Phys. Rev. Lett.},
  volume  = {121},
  pages   = {120401},
  year    = {2018},
  doi     = {10.1103/PhysRevLett.121.120401}
}

@article{Peppler2018,
  author  = {Peppler, Timo and Dyke, Paul and Zamorano, Michail and Herrera, Ilia and Hoinka, Sascha and Vale, Chris J.},
  title   = {Quantum Anomaly and 2D--3D Crossover in Strongly Interacting {F}ermi Gases},
  journal = {Phys. Rev. Lett.},
  volume  = {121},
  pages   = {120402},
  year    = {2018},
  doi     = {10.1103/PhysRevLett.121.120402}
}

@article{Murthy2019,
  author  = {Murthy, P. A. and Defenu, N. and Bayha, L. and Holten, M. and Preiss, P. M. and Jochim, S. and Enss, T.},
  title   = {Quantum scale anomaly and spatial coherence in a 2D {F}ermi superfluid},
  journal = {Science},
  volume  = {365},
  pages   = {268--272},
  year    = {2019},
  doi     = {10.1126/science.aau4402}
}

@article{Gasper1975,
  author  = {Gasper, George},
  title   = {Products of terminating ${}_3F_2(1)$ series},
  journal = {Pacific J. Math.},
  volume  = {56},
  pages   = {87--95},
  year    = {1975},
  doi     = {10.1016/0022-247X(75)90197-3},
}

@article{Stringari1996,
  author  = {Stringari, Sandro},
  title   = {Collective Excitations of a Trapped Bose-Condensed Gas},
  journal = {Phys. Rev. Lett.},
  volume  = {77},
  pages   = {2360--2363},
  year    = {1996},
  doi     = {10.1103/PhysRevLett.77.2360}
}

@article{Lobser2015,
  title={Observation of a persistent non-equilibrium state in cold atoms},
  author={Lobser, D. S. and Barentine, A. E. S. and Cornell, E. A. and Lewandowski, H. J.},
  journal={Nature physics},
  volume={11},
  number={12},
  pages={1009--1012},
  year={2015},
  publisher={Nature Publishing Group UK London},
  doi     = {10.1038/nphys3491},
}

@book{Szego1975,
  author    = {Szeg{\H{o}}, G{\'a}bor},
  title     = {Orthogonal Polynomials},
  series    = {American Mathematical Society Colloquium Publications},
  volume    = {23},
  edition   = {4th},
  publisher = {American Mathematical Society},
  year      = {1975}
}

@book{Lewin1981,
  author    = {Lewin, Leonard},
  title     = {Polylogarithms and Associated Functions},
  publisher = {North-Holland},
  year      = {1981}
}

@book{HTF1,
  author    = {Erd{\'e}lyi, A. and Magnus, W. and Oberhettinger, F. and Tricomi, F. G.},
  title     = {Higher Transcendental Functions, Vol. I},
  publisher = {McGraw--Hill},
  year      = {1953}
}

@article{Tan2008Energetics,
  author  = {Tan, Shina},
  title   = {Energetics of a strongly correlated {F}ermi gas},
  journal = {Annals of Physics},
  year    = {2008},
  volume  = {323},
  number  = {12},
  pages   = {2952--2970},
  month   = dec,
  doi     = {10.1016/j.aop.2008.03.004}
}

@article{Tan2008LargeK,
  author  = {Tan, Shina},
  title   = {Large momentum part of a strongly correlated {F}ermi gas},
  journal = {Annals of Physics},
  year    = {2008},
  volume  = {323},
  number  = {12},
  pages   = {2971--2986},
  month   = dec,
  doi     = {10.1016/j.aop.2008.03.005}
}

@article{Tan2008Virial,
  author  = {Tan, Shina},
  title   = {Generalized virial theorem and pressure relation for a strongly correlated {F}ermi gas},
  journal = {Annals of Physics},
  year    = {2008},
  volume  = {323},
  number  = {12},
  pages   = {2987--2990},
  month   = dec,
  doi     = {10.1016/j.aop.2008.03.003}
}

@article{BraatenPlatter2008OPE,
  author  = {Braaten, Eric and Platter, Lucas},
  title   = {Exact Relations for a Strongly Interacting {F}ermi Gas from the Operator Product Expansion},
  journal = {Physical Review Letters},
  year    = {2008},
  volume  = {100},
  pages   = {205301},
  doi     = {10.1103/PhysRevLett.100.205301}
}

@article{NishidaSon2007NRcft,
  author  = {Nishida, Yusuke and Son, Dam T.},
  title   = {Nonrelativistic conformal field theories},
  journal = {Physical Review D},
  year    = {2007},
  volume  = {76},
  pages   = {086004},
  doi     = {10.1103/PhysRevD.76.086004}
}

@article{KaganSurkovShlyapnikov1996PRA,
  author  = {Kagan, Yu. and Surkov, E. L. and Shlyapnikov, G. V.},
  title   = {Evolution of a Bose-condensed gas under variations of the confining potential},
  journal = {Physical Review A},
  year    = {1996},
  volume  = {54},
  number  = {3},
  pages   = {R1753--R1756},
  month   = sep,
  doi     = {10.1103/PhysRevA.54.R1753}
}

@article{KaganSurkovShlyapnikov1997PRA,
  author  = {Kagan, Yu. and Surkov, E. L. and Shlyapnikov, G. V.},
  title   = {Evolution of a Bose gas in anisotropic time-dependent traps},
  journal = {Physical Review A},
  year    = {1997},
  volume  = {55},
  number  = {1},
  pages   = {R18--R21},
  month   = jan,
  doi     = {10.1103/PhysRevA.55.R18}
}

@article{GueryOdelin1999PRA,
  author  = {Gu{\'e}ry-Odelin, David and Zambelli, Francesca and Dalibard, Jean and Stringari, Sandro},
  title   = {Collective oscillations of a classical gas confined in harmonic traps},
  journal = {Physical Review A},
  year    = {1999},
  volume  = {60},
  number  = {6},
  pages   = {4851--4856},
  month   = dec,
  doi     = {10.1103/PhysRevA.60.4851}
}

@article{DeSilva2008PRA,
  author  = {de Silva, Theja N.},
  title   = {Breathing mode frequencies of a rotating {F}ermi gas in the {BCS}-{BEC} crossover region},
  journal = {Physical Review A},
  year    = {2008},
  volume  = {78},
  pages   = {023623},
  month   = aug,
  doi     = {10.1103/PhysRevA.78.023623}
}

@article{Son2007PRL_BulkViscosity,
  author  = {Son, D. T.},
  title   = {Vanishing Bulk Viscosities and Conformal Invariance of the Unitary {F}ermi Gas},
  journal = {Physical Review Letters},
  year    = {2007},
  volume  = {98},
  pages   = {020604},
  doi     = {10.1103/PhysRevLett.98.020604}
}

@article{Enss2019PRL_BulkViscosityContact,
  author  = {Enss, Tilman},
  title   = {Bulk Viscosity and Contact Correlations in Attractive {F}ermi Gases},
  journal = {Physical Review Letters},
  year    = {2019},
  volume  = {123},
  pages   = {205301},
  doi     = {10.1103/PhysRevLett.123.205301}
}

@article{Mulkerin2018_PRA_collective2D,
  author  = {Brendan C. Mulkerin and Xia-Ji Liu and Hui Hu},
  title   = {Collective modes of a two-dimensional {F}ermi gas at finite temperature},
  journal = {Physical Review A},
  volume  = {97},
  number  = {5},
  pages   = {053612},
  year    = {2018},
  doi     = {10.1103/PhysRevA.97.053612}
}

@article{Yin2020_PRL_fewbodyAnomaly2D,
  author  = {Xiang-Yu Yin and Hui Hu and Xia-Ji Liu},
  title   = {Few-Body Perspective of a Quantum Anomaly in Two-Dimensional {F}ermi Gases},
  journal = {Physical Review Letters},
  volume  = {124},
  number  = {1},
  pages   = {013401},
  year    = {2020},
  doi     = {10.1103/PhysRevLett.124.013401}
}

@article{BekassyHofmann2022_PRL_conformalMesoscopic2D,
  author  = {Viktor Bekassy and Johannes Hofmann},
  title   = {Nonrelativistic Conformal Invariance in Mesoscopic Two-Dimensional {F}ermi Gases},
  journal = {Physical Review Letters},
  volume  = {128},
  number  = {19},
  pages   = {193401},
  year    = {2022},
  doi     = {10.1103/PhysRevLett.128.193401}
}

@article{Toniolo2018_PRA_crossoverBreathing,
  author  = {Umberto Toniolo and Brendan C. Mulkerin and Xia-Ji Liu and Hui Hu},
  title   = {Breathing-mode frequency of a strongly interacting {F}ermi gas across the two- to three-dimensional crossover},
  journal = {Physical Review A},
  volume  = {97},
  number  = {6},
  pages   = {063622},
  year    = {2018},
  doi     = {10.1103/PhysRevA.97.063622}
}

@article{Chiacchiera2013_PRA_quadDamping2D,
  author  = {Silvia Chiacchiera and Dany Davesne and Tilman Enss and Michael Urban},
  title   = {Damping of the quadrupole mode in a two-dimensional {F}ermi gas},
  journal = {Physical Review A},
  volume  = {88},
  number  = {5},
  pages   = {053616},
  year    = {2013},
  doi     = {10.1103/PhysRevA.88.053616}
}

@article{gao2012breathing,
  title={Breathing mode of two-dimensional atomic {F}ermi gases in harmonic traps},
  author={Gao, Chao and Yu, Zhenhua},
  journal={Physical Review A},
  volume={86},
  number={4},
  pages={043609},
  year={2012},
  publisher={APS},
  doi={10.1103/PhysRevA.86.043609}
}

@article{hu2019reduced,
  title={Reduced quantum anomaly in a quasi-two-dimensional {F}ermi superfluid: significance of the confinement-induced effective range of interactions},
  author={Hu, Hui and Mulkerin, Brendan C. and Toniolo, Umberto and He, Lianyi and Liu, Xia-Ji},
  journal={Physical Review Letters},
  volume={122},
  number={7},
  pages={070401},
  year={2019},
  publisher={APS},
  doi={10.1103/PhysRevLett.122.070401}
}

@article{He2020,
  author  = {Chengdong He and Zejian Ren and Bo Song and Entong Zhao and Jeongwon Lee and Yi-Cai Zhang and Shizhong Zhang and Gyu-Boong Jo},
  title   = {Collective excitations in two-dimensional SU({N}) {F}ermi gases with tunable spin},
  journal = {Physical Review Research},
  volume  = {2},
  number  = {1},
  pages   = {012028},
  year    = {2020},
  doi     = {10.1103/PhysRevResearch.2.012028}
}

@article{Bohlen2020,
  author  = {Markus Bohlen and Lennart Sobirey and Niclas Luick and Hauke Biss and Tilman Enss and Thomas Lompe and Henning Moritz},
  title   = {Sound Propagation and Quantum-Limited Damping in a Two-Dimensional {F}ermi Gas},
  journal = {Physical Review Letters},
  volume  = {124},
  number  = {24},
  pages   = {240403},
  year    = {2020},
  doi     = {10.1103/PhysRevLett.124.240403}
}

@article{LippariniStringari1989,
  author    = {E. Lipparini and S. Stringari},
  title     = {Sum rules and giant resonances in nuclei},
  journal   = {Physics Reports},
  volume    = {175},
  pages     = {103--261},
  year      = {1989},
  doi       = {10.1016/0370-1573(89)90029-X},
}

@article{Olshanii2010,
  author  = {Maxim Olshanii and H\'el\`ene Perrin and Vincent Lorent},
  title   = {Example of a Quantum Anomaly in the Physics of Ultracold Gases},
  journal = {Phys. Rev. Lett.},
  volume  = {105},
  pages   = {095302},
  year    = {2010},
  doi     = {10.1103/PhysRevLett.105.095302},
}

@article{Moroz2012,
  author  = {Sergej Moroz},
  title   = {Scale-invariant {F}ermi gas in a time-dependent harmonic potential},
  journal = {Phys. Rev. A},
  volume  = {86},
  pages   = {011601},
  year    = {2012},
  doi     = {10.1103/PhysRevA.86.011601},
}

@article{Dalfovo1999,
  author  = {Dalfovo, Franco and Giorgini, Stefano and Pitaevskii, Lev P. and Stringari, Sandro},
  title   = {Theory of {{B}ose-{E}instein} condensation in trapped gases},
  journal = {Rev. Mod. Phys.},
  volume  = {71},
  pages   = {463--512},
  year    = {1999},
  doi     = {10.1103/RevModPhys.71.463},
}

@article{Serwane2011Science,
  author  = {Serwane, F. and Z\"urn, G. and Lompe, T. and Ottenstein, T. B. and Wenz, A. N. and Jochim, S.},
  title   = {Deterministic Preparation of a Tunable Few-{F}ermion System},
  journal = {Science},
  volume  = {332},
  pages   = {336--338},
  year    = {2011},
  doi     = {10.1126/science.1201351},
}

@article{Zurn2012PRL,
  author  = {Z\"urn, G. and Serwane, F. and Lompe, T. and Wenz, A. N. and Ries, M. G. and Bohn, J. E. and Jochim, S.},
  title   = {{F}ermionization of Two Distinguishable {F}ermions},
  journal = {Phys. Rev. Lett.},
  volume  = {108},
  pages   = {075303},
  year    = {2012},
  doi     = {10.1103/PhysRevLett.108.075303},
}

@article{Hood2020PRResearch,
  author  = {Hood, J. D. and Yu, Y. and Lin, C. and Caldwell, R. and Gavryusev, V. and Raizen, M. G.},
  title   = {Multichannel interactions of two atoms in an optical tweezer},
  journal = {Phys. Rev. Research},
  volume  = {2},
  pages   = {023108},
  year    = {2020},
  doi     = {10.1103/PhysRevResearch.2.023108},
}

\end{document}